\documentclass[10pt,aps,twocolumn]{revtex4}
\usepackage{graphicx}
\usepackage{amssymb}
\usepackage{epsfig}
\usepackage{bm}
\usepackage{amsmath,graphicx,subfigure}

\newcommand{\vecc}[1]{\mbox{\boldmath $#1$}}
\newcommand\nn{\nonumber}
\newcommand\dd{\mathrm{d}}

\def\({\left(}
\def\[{\left[}
\def\){\right)}
\def\]{\right]}
\def\a{\alpha}
\def\b{\beta}

\begin{document}

\title{Double-logarithmic behavior of inelastic fermion form factors
in QED and QCD}

\author{E.~Barto\v{s}}\email{bartos@thsun1.jinr.ru}
\altaffiliation{On leave of absence from the Department of
Theoretical Physics, Comenius University, 84248 Bratislava,
Slovakia.}
\author{E.~A.~Kuraev}\email{kuraev@thsun1.jinr.ru}
\author{I.~O.~Cherednikov}\email{igor.cherednikov@jinr.ru}
\affiliation{Joint Institute for Nuclear Research \\
RU-141980 BLTP JINR, Dubna, Russia}


\begin{abstract}
The effective kinematic diagram technique is applied to study
inelastic form factors of electron and quark in QED and QCD. The
explicit expressions for these form factors in the
double-logarithmic approximation are presented. The
self-consistency of the results is shown by demonstrating the
fulfillment of the Kinoshita-Lee-Nauenberg theorem.
\end{abstract}

\maketitle

\section{Introduction}

Precise computation of the elastic and inelastic fermion form
factors in hard collisions is required to test the predictive
power of the Standard Model, as well as the effective and
unambiguous detection of signals of New Physics at modern and
future colliders (see, {\it e.g.}, \cite{COLL} and references
therein). In QCD, the quark form factors are used in calculations
of various QCD processes at the partonic level, and are of a
considerable phenomenological importance \cite{HARD}. In
investigation of $e^+e^-$ collisions at $TeV$ energies, the
resumed leading and nonleading Sudakov corrections to fermion form
factors may profoundly influence the cross sections, and play a
significant role in calculations for the Next Linear Collider
\cite{KPS, FLMM, JP}.

Since the pioneering calculation of the resumed double-logarithmic
(DL) corrections to the elastic electron form factor in QED
\cite{Sudakov}, significant progress has been made in evaluation
of the next-to-leading logarithmic contributions \cite{BMR,
ALLOG}, as well as in generalization of these results to the
strong (QCD) and electroweak (EW) sectors of the Standard Model
(see, {\it e.g.}, \cite{KPS, ET, FLMM}, and references therein).

The well-known Sudakov elastic form factor $F(q^2)$ of the
electron scattering in external electromagnetic field with large
transferred momentum $q = p_2-p_1$
\begin{equation}\label{}
\mathrm{e}(p_1)+\gamma^*(q)\to{\mathrm e}(p_2)
\end{equation}
has the form \cite{Sudakov}
\begin{equation}\label{}
F(s)=\exp\(-\frac{\a}{4\pi} \ \ln^2 \frac{s}{\lambda^2}\) \ , \
s\gg \lambda^2 \ ,
\end{equation}
where $s= -q^2$ and the mass $\lambda$ of the virtual vector boson
is introduced in order to regulate the IR divergence.

Such a strong suppression of elastic form factor is quite natural
since it reflects a small probability for an electron to remain to
be itself in this process. Therefore, inelastic processes with
emission of one or several real vector bosons become more
probable. Although all the exclusive scattering probabilities
experience the Sudakov type suppression, the total sum of them
must be equal to 1 and possess no singularities in the massless
limit in accordance with the Kinoshita-Lee-Nauenberg (KLN)
theorem. Such a cancellation can be easily proven in QED using the
Poisson nature of the inelastic form factor. As for QCD, the
problem of KLN cancellation is more complicated due to violation
of the Poisson form of form factors. Nevertheless, the explicit
expressions for inelastic form factors with radiative corrections
taken into account can be obtained in certain kinematics of the
real gluon emission which can be realized in experiment. An
effective way to get them is to apply the kinematic diagram
method. It is the motivation of our paper to calculate the
inelastic fermion form factors within this framework.

\section{Description of the method}

In this paper, we derive the inelastic form factor of a fermion
(electron, or quark) in the double-logarithmic (DL) approximation
\begin{eqnarray}
g^2\ll 1 \ ,\quad g^2L^2\gg 1 \ ,\quad
L=\ln\frac{s}{\lambda^2},\quad s=-q^2\gg \lambda^2 \ ,
\end{eqnarray}
where $\lambda$ is the mass of a virtual vector boson, and
$g^2=4\pi\a$. A powerful method for calculation of the elastic
cross sections in this approximation was developed in the Quantum
ElectroDynamics (QED) and in the Quantum ChromoDynamics (QCD) (see
\cite{KL} and references therein). It was found out that the
straightforward calculation of the Feynman diagrams was not the
most economical way to resume the DL asymptotics of form factors
and cross sections. Here we apply a different approach based on
the use of the effective kinematic diagrams.

Throughout the paper we use the Sudakov representation of the
four-momenta of the virtual bosons (neutral massive vector
particles, or massive gluons)
\begin{eqnarray}
k=\a p_2 + \b p_1 + k_\bot \ ,
\end{eqnarray}
where $p_1$, $p_2$ are the four-momenta of the external fermions,
and  $q=p_2-p_1$ is the transferred momentum in the scattering
channel. Let us remind the main features of the Sudakov
parameterization
\begin{gather}
k_\bot p_1=k_\bot p_2=0 \ ,\quad k^2=s\alpha\beta-\vecc{k}^2 \ ,\\
\nn k_{i\bot}k_{j\bot}=-\vecc{k}_i\vecc{k}_j \ ,\quad
\dd^4k=\frac{s}{2}\:\dd\alpha\: \dd\beta\: \dd^2\vecc{k} \ .
\end{gather}

The ground of the effective kinematic diagram method is twofold.
The first reason is the strong ordering of the virtual photons in
magnitude of transversal components of their four-momenta. For a
set of Feynman diagrams (FD) with $n$ virtual vector bosons, the
main DL contribution arises from the region where their
transversal momenta are strictly ordered
\begin{eqnarray}
s\gg\vecc{k}_{i_1}^2\gg\vecc{k}_{i_2}^2\gg\dots\gg
\vecc{k}_{i_n}^2\gg \lambda^2 \ .
\end{eqnarray}
The second reason is the Gribov theorem about validity of the
classical current approximation for the emission of vector bosons
in the extended region \cite{GRIBOV}. The main idea of this
theorem (in a particular case of the DL calculations) can be
expressed in the following manner: The amplitude of the process $
q(p)+g(k) \to X$ can be related with the amplitude of the
transition of the almost on-mass-shell quark $q(p)$ to the same
state $X$
\begin{gather}
{\cal M}\( q(p)+g(k) \to X \)=\frac{\epsilon \cdot p }{k\cdot p}
{\cal M}\( q(p^*)\to X \) \ ,
\end{gather}
if the transversal component of its 4-momentum is small in
comparison with the characteristic transversal momentum in the
block $X$, where $\epsilon$ is the polarization vector of the
gluon and $(p^*)^2 \approx m^2$. Thus, the Gribov theorem refutes
the common belief that the classical current approximation is
valid only for soft photons.

Consider now the interaction of a virtual gluon having minimal
transversal momentum $\vecc{k_m}$ with a quark of momentum $p_1$
(Fig.~(1a)). The corresponding amplitude being an analytical
function of the variable $(p+k_m)^2$ has the pole corresponding to
the one-quark intermediate state and the cut which corresponds to
the quark-gluon intermediate state (Fig.~(1b)). Let us now prove
that the contribution of the cut is suppressed in the DL
approximation while it can contribute to the nonleading terms
which do not survive in the asymptotic regime. Indeed, keeping in
the mind the current conservation condition for the amplitude of
the block $g + q \to X$ \cite{KL, GRIBOV}
$$ k^\mu {\cal M}^X_\mu = \( \a p_2 + k_\bot \)^\mu {\cal M}^X_\mu =0 \ , $$ and the
Green function of a vector boson with momentum $k$:
\begin{equation}
G_{\mu\nu}=\frac{g^\bot_{\mu\nu}+
{\frac{2}{s}}\(p_{1\mu}p_{2\nu}+p_{2\mu}p_{1\nu}\)}{\(k^2 -
\lambda^2 + i0\) } \ ,
\end{equation}
(there are no ghosts in this gauge) one finds that the cut
contribution is associated with the additional factor depending on
$|\vecc{k}|/|\vecc{k}_i|\ll 1$ compared with the contribution of
the pole FD.

Therefore, in the DL approximation the softest virtual photon
(gluon) effectively interacts with the quarks having the momenta
$p_1$ and $p_2$, {\it i.e.}, it is emitted before all the photons
(gluons) counting along the quark line, and absorbed after all
other gluons. Similar reasons lead to construction of a ladder
type FD with all the rungs parallel to each other. In calculation
of the corresponding amplitude it is implied that the virtualities
of the vector bosons are strictly ordered (Fig.~(1c)).
\begin{figure}[tb] \label{fig:1}
\subfigure[] {\includegraphics[scale=.7]{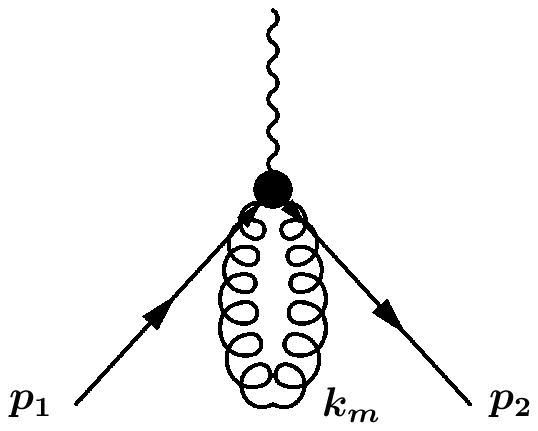}} \subfigure[]
{\includegraphics[scale=.55]{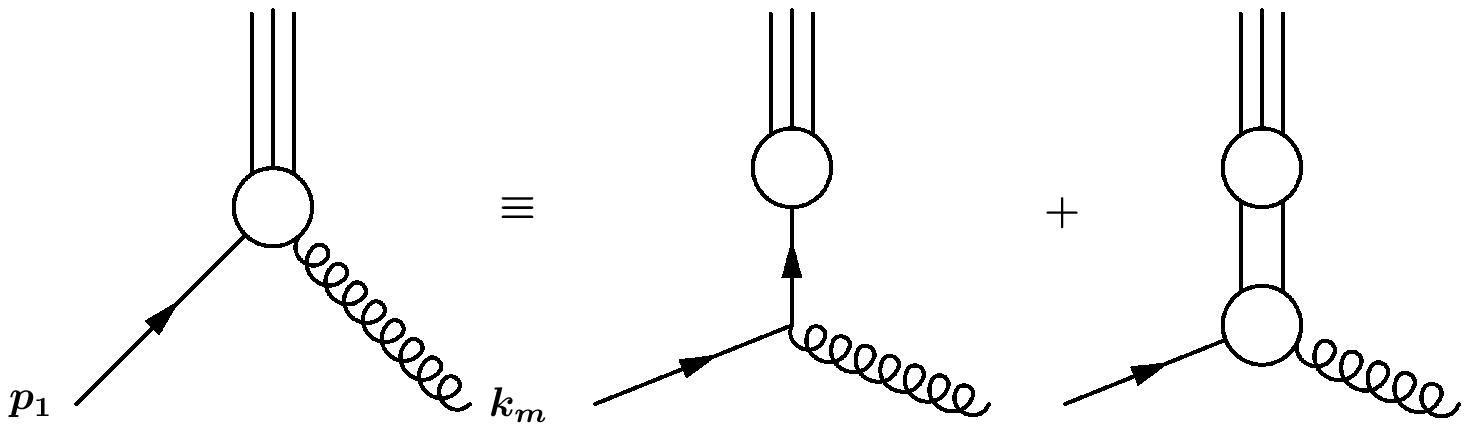}} \subfigure[]
{\includegraphics[scale=.7]{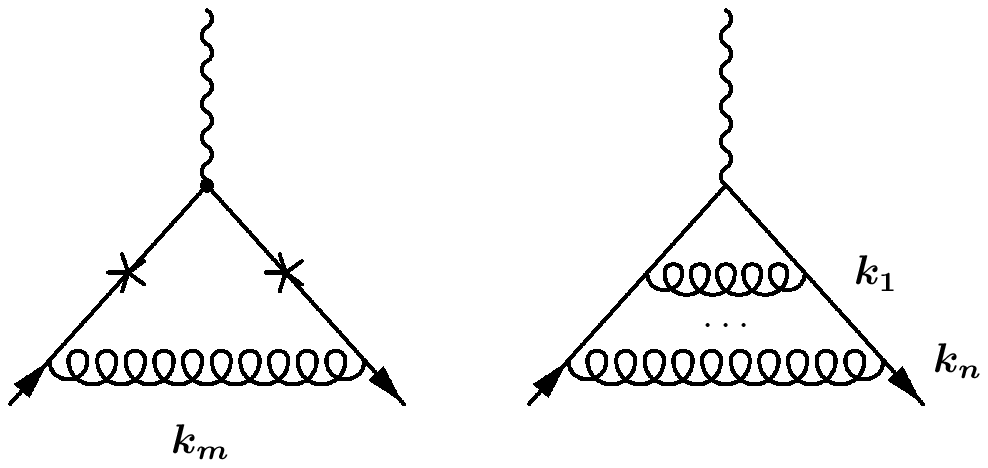}} \caption{The kinematic
diagrams for (a) the extracted gluon $g_{k_m}$ with the minimal
value of the transversal momentum $\vecc k_m$, (b) the pole and
cut contributions to the quark-gluon amplitude, (c) the pole
dominant kinematic diagrams with descending momenta
$\vecc{k}_1^2\gg \vecc{k}_2^2\gg\dots\gg \vecc{k}_n^2$.}
\end{figure}

It is easy to understand that due to this ordering the
contributions of such FD can be expressed in terms of the lowest
order (Born) amplitude $B^{(0)}$ as ${B^{(0)}}^n/n!$, which leads
to the Sudakov type form factor
\begin{eqnarray}
F^{(0)}(s)=\mathrm{e}^{- B^{(0)}} \ .
\end{eqnarray} The condition of ordering can be removed
when one considers the whole set of $n!$ similar expressions
obtained by symmetrization of the momenta indices. Thus, the
combinatorial factor $1/n!$ must be introduced.

This type of the DL behavior can be obtained by explicit
calculations in lowest orders of PT in both QED \cite{Akhiez},
$B^{(0)}_{\mathrm{QED}}=\(e^2/16\pi^2\) \ L^2$, and QCD
\cite{Tiktop}, $B^{(0)}_{\mathrm{QCD}}=\( \a_s/4\pi \) C_F L^2$.
In Appendix we derive these lowest order expressions.

It can be shown that a possible contribution of longitudinal
polarized virtual and real vector particles is suppressed due to
the gauge invariance and is irrelevant in the DL regime.

\section{Inelastic form factors for one vector boson emission}

\begin{figure}[tb] \label{fig:2}
\centering \subfigure[]
{\includegraphics[scale=.5]{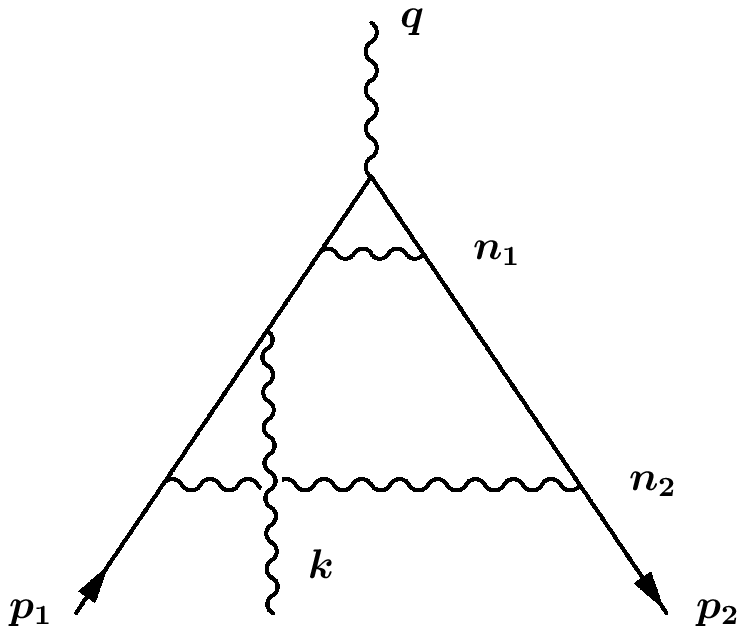}}\hspace{3mm} \subfigure[]
{\includegraphics[scale=.5]{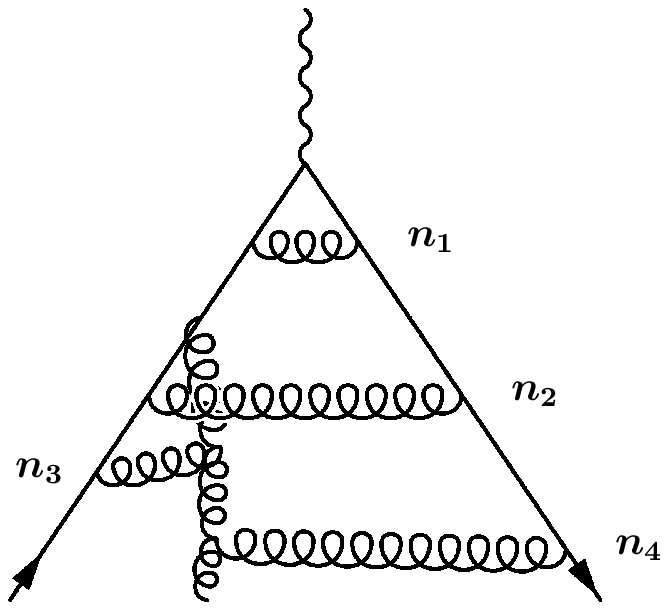}}\caption{The QED kinematic
diagram for the hard photon emission (a) and the QCD kinematic
diagram for the hard gluon emission (b).}
\end{figure}

Now let us consider the inelastic electron form factor which
includes the emission of a photon with momentum $k_1$ and
polarization vector $\epsilon(k_1)$ (Fig.~(2a)). We consider the
situation when the transversal momentum of this real hard photon
is large compared to the virtual photon mass $\lambda$ and the
masses of fermions:
\begin{eqnarray}
\vecc{k}_1^2=s\a_1\b_1 \gg \lambda^2, m^2 \ .
\end{eqnarray}
This corresponds to the kinematics which produces the main
contribution to the total cross section. Indeed, consider for
estimation the contribution to the total cross section in the
classical current approximation:
\begin{gather}
-\int\frac{\dd^3k_1}{\omega_1}j^2(k_1) \sim\int\limits_0^{\pm1}
\int\limits_0^{\pm1}\frac{\dd\a_1}{\a_1}\frac{\dd\b_1} {\b_1}
\theta(s\a_1\b_1-\lambda^2)= L^2 \ ,\\ \nn j^\mu(k_1)=\(
\frac{p_1}{p_1k_1}-\frac{p_2}{p_2k_1}\)^\mu \ .
\end{gather}

Again, the effective kinematic ladder FD approach can be applied,
but the large magnitude of $\vecc{k}_1^2$ requires some
modification in the ordering procedure. Namely, for the virtual
photons emitted above the point where the real photon is emitted
(which is closer to the point of interaction with the external
particle) we must choose the quantity $\vecc{k}^2_1$ as a lower
bound for $\vecc{k}^2_i$ . The virtual photons emitted below the
point of the external photon emission have $\vecc{k}_1^2$ as an
upper bound. Therefore, the restriction has the following form
$\lambda^2 \ll \vecc{k}_j^2\ll\vecc{k}^2_1$. Denoting the number
of the ``up'' photons by $n_1$ and the ``down'' ones by $n-n_1$,
one obtains the contribution of $n$ virtual photons to the
amplitude of the one-photon radiative scattering
\begin{widetext}
\begin{equation}
{\cal M}^{(1)}_n =  eV_0j^\mu(k_1) \epsilon_\mu(k_1)\(-\frac{e^2}
{16\pi^2} \)^n \sum_{n_1=0}^{n}
\frac{(L_1^2)^{n_1}}{(n_1)!}\frac{\(L^2-L_1^2\)^{n-n_1}}{(n-n_1)!}
=e V_0 j^\mu(k_1) \epsilon_\mu(k_1)\frac{\(-\frac{e^2
L^2}{16\pi^2}\)^n}{n!} \ , \label{em1ph}
\end{equation}
where
\begin{gather}
L_1=\ln\frac{s}{\vecc{k}_1^2} \ ,\quad V_0=e\bar{u}(p_2)\Gamma
u(p_1) \ ,\quad
\Gamma=(1;\gamma_5;\gamma_\rho;\gamma_5\gamma_\rho) \ . \nn
\end{gather}
\end{widetext}
Further resummation is straightforward.

In the case of emission of $k$ real hard photons we have
\begin{eqnarray}
{\cal M}^{(k)}_\infty = e^k\
V_0\prod_{j=1}^{k}j^\mu(k_j)\epsilon_\mu(k_j)\mathrm{e}^{-B^{(0)}/2}
\  .
\end{eqnarray}
It is implied that a hard photon is a photon with transversal
momentum much larger than masses of fermions and virtual photons.
The contribution to the total cross section is associated with the
factor
\begin{eqnarray} \label{eq:pois}
F_\infty^{(k)}=\frac{{B^{(0)}}^k}{k!}\mathrm{e}^{-B^{(0)}} \ ,
\end{eqnarray}
confirming the Poisson nature of the neutral vector bosons
emission. The factor $1/k!$ takes into account the identity of the
emitted bosons. Therefore, we can see that the Poisson
distribution is valid not only for the soft photons \cite{Akhiez},
but also for the hard ones in the DL approximation \cite{GOR}.

In order to establish the consistency of the result we verify the
fulfillment of the KLN theorem \cite{KLN} about cancellation of
the mass singularities, namely
\begin{equation}
\sum_{n=0}^\infty \ F_\infty^{(n)} = 1  \ .
\end{equation}

In QCD, the ladder approach for calculation of the inelastic form
factors with emission of a single gluon works as well. However,
now the virtual ladder gluons can interact (besides the quarks)
with one real hard gluon with momentum $k_1$. Let us show that in
this case the inelastic form factor has the form
\begin{gather} \label{eq:ef}
F_\infty^{(1)}=F_0^{(1)}\exp\[-\frac{\a_s}{4\pi} \(C_F \ L^2 +
\frac{C_V}{2} L_k^2\) \]\ , \\ \nn F_0^{(1)}=g \ V_{0b}\
j^\mu(k_1)\epsilon_\mu(k_1)\ ,\quad
L_k=\ln\frac{\vecc{k}_1^2}{\lambda^2} \  .
\end{gather}
where $V_{0b}=\bar{u}(p_2)\sigma_b\Gamma u(p_1)$ is the
corresponding Born amplitude, $C_F=\frac{N_c^2-1}{2N_c}$,
$C_V=N_c$ are the Casimir operators of the color group $SU(N_c)$,
and $\sigma_b$'s are the group generators. The eight kinematic FD
exist in the one-loop order. It is sufficient to consider only
four of them which describe the emission from the quark 1 (for
simplicity, we denote the quark with the momentum $p_{1,2}$ by
``quark 1,2''). The color factor associated with the region when
the loop momentum $|\vecc{k}|$ is large as compared with
$|\vecc{k}_1|$ is $\sigma_a\sigma_a=C_F \ I$, since the color
generators commute with the external vertex operator $\Gamma$.
This gives the contribution similar to the QED case:
\begin{eqnarray} \label{eq:13}
-Z_1  C_F  L_k^2, \quad Z_1=g V_{0b}\frac{p_1\cdot
\epsilon_1}{p_1\cdot k_1}\frac{\a_s}{4\pi} \ , \quad
\epsilon_1=\epsilon(k_1) \ .
\end{eqnarray}
The contribution of another QED-type FD corresponding to the case
$|\vecc{k}_1|\gg|\vecc{k}|$ is accompanied by the color factor
$\sigma_a\sigma_b\sigma_a = (C_F-C_V/2)\sigma_b$. Its contribution
reads
\begin{eqnarray}
-Z_1\(C_F-\frac{C_V}{2}\) \(L^2-L_k^2\) \ .
\end{eqnarray}
The color factor of the FD that corresponds to the case
$|\vecc{k}_1|\gg|\vecc{k}|$, when the gluon is emitted by the
quark 1 and absorbed by the external gluon, is $i
f_{abc}\sigma_a\sigma_c=C_V/2$. The corresponding kinematic
contribution without the restriction imposed by the real gluon
emission is given by
$$\int\limits_{\frac{\lambda^2}{s}}^{\alpha_1}\frac{\dd\alpha}{\alpha}
\int\limits_{\frac{\lambda^2}{s\alpha}}^1\frac{\dd\beta}{\beta} \
.$$ The restriction consists in the subtraction of a similar
expression with replacement $\lambda^2\to\vecc{k}^2_1$. The total
contribution of this FD reads
\begin{eqnarray}
-Z_1\frac{1}{2} C_V \[\ln^2\frac{s\alpha_1}{\lambda^2}-
\ln^2\frac{s\alpha_1}{\vecc{k}_1^2}\].
\end{eqnarray}
The contribution of the FD with the gluon rung connecting  the
real gluon with the quark 2 can be obtained from the last
expression by means of the replacement $\a_1 \to \b_1$.

The total one-loop contribution to the one-gluon radiative
scattering of a quark  (including the FD with the real gluon
emitted by the quark 2) has the form
\begin{multline}
{\cal M}^{(1)}_1  = gV_{0b}j^\mu(k_1)\epsilon_\mu(k_1) \\ \cdot \[
-\frac{\a_s}{4\pi} \(C_F \ L^2+\frac{1}{2} C_V \ \(L-L_1\)^2 \) \]
\ .
\end{multline}
This result agrees with that one obtained in the work by one of us
(see Eq. (11) in the Ref. \cite{KF}).

The FD with $n$ loops can be parameterized by the numbers $n_1$,
$n_2$, $n_3$, $n_4$ (Fig.~(2b)) which are, respectively,
$n_1$---the number of rungs with $\vecc{k}_i^2\gg\vecc{k}_1^2$;
$n_2$---the number of rungs with $\vecc{k}_1^2\gg\vecc{k}_i^2$
connecting the quarks, and $n_3$---the quark 1 with the real gluon
; ($n_4$)---the number of rungs connecting the real gluon with the
quark 2, and $n=n_1+n_2+n_3+n_4$. For the real gluon emitted by
the quark 1 one has
\begin{multline}
(-1)^n\sum\frac{1}{n_1!}\(C_F \ L_k^2\)^{n_1}\[\(C_F-
\frac{1}{2}C_V\)\(L^2-L_k^2\)\]^{n_2} \\ \cdot
\frac{1}{n_3!}\(\frac{1}{2}C_V\ln\frac{\vecc{k}_1^2}{\lambda^2}
\ln\frac{s\alpha_1}{\beta_1\lambda^2}\)^{n_3}\\ \cdot
\frac{1}{n_4!}\(\frac{1}{2}C_V\ln\frac{\vecc{k}_1^2}{\lambda^2}
\ln\frac{s\beta_1}{\alpha_1\lambda^2}\)^{n_4}\\
=(-1)^n\frac{1}{n!}\[C_F \ L^2+\frac{1}{2} C_V\ \(L-L_1\)^2\]^n \
,
\end{multline}
confirming the relation given above (see Eq. (\ref{eq:ef})).

\section{Inelastic form factors for arbitrary number of
emitted vector bosons}
\begin{figure}[tb] \label{fig:3}
\centering
\includegraphics[scale=.7]{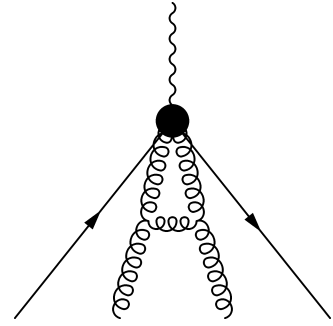}
\caption{The kinematic diagram yielding the color exotic
contribution.}
\end{figure}

Consider now the emission of two hard gluons. Longitudinal Sudakov
parameters of the gluon momenta $k_i=\alpha_i p_2+\beta_i
p_1+k_{i\bot}$ in the region of main contribution to the cross
section obey the following restrictions
\begin{eqnarray} \frac{\lambda^2}{s}\sim
\a_1 \ll \a_2\sim 1 \ ,\quad \frac{\lambda^2}{s}\sim
\b_2\ll\b_1\sim 1 \ .
\end{eqnarray}
The corresponding Born amplitude has the form
\begin{multline}
g^2\[ V_{0ab}(\a_2\b_1\gg\a_1\b_2) + V_{0ba} (\a_1\b_2 \gg
\a_2\b_1) \] \\ \cdot j^\mu(k_1)\epsilon_\mu(k_1)
j^\nu(k_2)\epsilon_\nu(k_2) \ ,
\end{multline}
with $V_{0ab}=\bar{u}(p_2) \Gamma \sigma_a\sigma_b u(p_1)$. Note,
that we do not consider here the region $\a_2\b_1=\a_1\b_2$ which
also yields the DL contributions to the amplitude as it was shown
by rather complicated calculations given in \cite{KF}. This
kinematic region is specific of QCD and corresponds to the decay
of a gluon to two gluons. In \cite{EF, Ermol}, the arguments in
favour of exponentiation of the DL contributions, including also
the decay mechanism, were given.

Consider the emission of both gluons from the quark 1 leg (the
similar situation takes place for any other kinematic regions of
the two-gluon emission). We have (here and below we use the
notation $\sigma_{ij}=\sigma_i\sigma_j$)
\begin{eqnarray}
\frac{p_1\cdot \epsilon_1}{p_1 \cdot k_1}\frac{p_1 \cdot
\epsilon_2}{p_1 \cdot k_2}
\[
\sigma_{21}\Bigg|^{\a_2\b_1 \gg \a_1\b_2}_{\a_1\b_1 \gg \a_2\b_2}
+ \sigma_{12}\Bigg|^{\a_1\b_2 \gg \a_2 \b_1}_{\a_2\b_2 \gg
\a_1\b_1}
\] \ .
\end{eqnarray}

For the 1-loop radiative corrections (RC) to this process one
needs to distinguish 3 kinematic regions
\begin{gather}
\vecc{q}^2\gg \vecc{k}^2\gg \vecc{k}_2^2 \ ,\quad \vecc{k}_2^2\gg
\vecc{k}^2\gg \vecc{k}_1^2 \ ,\quad \vecc{k}_1^2\gg \vecc{k}^2\gg
\lambda^2 \ .
\end{gather}

There are 10 kinematic FD of that type yielding the contribution
\begin{multline}
C_F\sigma_{21}L_{k_2}^2+\(C_F - \frac{C_V}{2}\)\sigma_{21}
\(L_{k_1}^2-L_{k_2}^2\)\\
+\frac{C_V}{2}\sigma_{21}\(L_{k_1}^2-L_{k_2}L_{k_1}\)+
\sigma_a\sigma_{21}\sigma_a\(L^2-L{k_1}^2\)\\
+\frac{C_V}{2}\sigma_{21}
\(\ln^2\frac{s\a_1}{\lambda^2}-\ln^2\frac{s\a_1}
{\vecc{k}_1^2}\)\\+
\frac{C_V}{2}\sigma_{21}\(\ln^2\frac{s\b_2}{\lambda^2}-
\ln^2\frac{s\b_2}{\vecc{k}_1^2}\)\\+
f_{k2b}f_{k1a}\sigma_b\sigma_a\(\ln^2\frac{2k_1k_2}{\lambda^2}-
\ln^2\frac{2k_1k_2}{\vecc{k}_1^2}\)\\
-if_{ba1}\sigma_b\sigma_2\sigma_a\(\ln^2\frac{s\b_1}{\lambda^2}-
\ln^2\frac{s\b_1}{\vecc{k}_1^2}\)\\
-if_{ba2}\sigma_b\sigma_1\sigma_a\(\ln^2\frac{s\a_2}{\lambda^2}-
\ln^2\frac{s\a_2}{\vecc{k}_1^2}\) \ .
\end{multline}
Rearranging the color indices using the relations
$[\sigma_a,\sigma_b]=if_{abc}\sigma_c$ and
$f_{abk}f_{abm}=C_V\delta_{km}$, one can see that the factor
accompanying the new color (exotic) structure
$f_{a2b}f_{a1k}\sigma_b\sigma_k$ is exactly equal to zero
(Fig.~(3)). The cancellation of such a kind of exotics takes place
in the higher orders of PT as well.

Hence, the result for the amplitude with two emitted gluons reads
\begin{equation}\label{}
{\cal M}^{(2)} = F_0^{(2)}B_{\mathrm{QCD}}^{(2)} \ ,
\end{equation}
where
\begin{multline}
F_0^{(2)}=\frac{p_1 \cdot \epsilon_1}{p_1\cdot k_1}\frac{p_1 \cdot
\epsilon_2}
{p_1 \cdot k_2}g^2 \ , \\
B_{\mathrm{QCD}}^{(2)}=\frac{\a_s}{2\pi} \[C_F \ L^2 +
\frac{C_V}{2} \( \ln^2 \frac{\vecc{k}_1^2}{\lambda^2}+ \ln^2
\frac{\vecc{k}_2^2}{\lambda^2}\)\] \ .
\end{multline}
Arguments in favour of exponentiation of higher orders of PT allow
one to conclude that
\begin{equation}\label{}
{\cal M}_\infty^{(2)}=F_0^{(2)}\exp\[- B_{\mathrm{QCD}}^{(2)}\] \
.
\end{equation}

Let us give some reasons for the following form of inelastic quark
form factor with emission of $m$ real hard gluons:
\begin{widetext}
\begin{equation} \label{eq:be}
{\cal M}_\infty^{(m)}={\cal M}_0^{(m)}\exp\[-\frac{\a_s}{4\pi}\(
C_F \ L^2 + \frac{C_V}{2} \sum_{i=1}^{m}\ln^2
\frac{\vecc{k}_i^2}{\lambda^2} \)\] \ ,
\end{equation}
where the amplitudes $B_0^{(m)}$ in the Born approximation are
given by
\begin{equation}
B_0^{(m)}=g^m\prod_{i=1}^m \epsilon_\mu(k_i)
j_\mu(k_i)\sum_{perm}\bar{u}(p_2)\Gamma\sigma_{a_1}...
\sigma_{a_m} u(p_1) \ ,
\end{equation}
and the following ordering takes place
\begin{gather} \label{eq:kin}
\a_{a_1}\gg\a_{a_2}\gg\dots\gg\a_{a_n} \ ,\quad
\b_{a_1}\ll\b_{a_2}\ll\dots\ll\b_{a_n}\ , \quad
s\a_{a_i}\b_{a_i}=\vecc{k_i}^2 \ .
\end{gather}
\end{widetext}
The kinematic conditions (\ref{eq:kin}) provide the extraction of
the leading DL contributions to the integrated hard gluon
distribution.

Consider now the ladder amplitude ${\cal M}_n^{(m)}$ with $m+1$
rungs, each of which with $n_j$ virtual gluons $(\sum_{j=0}^m
n_j=n)$. According to magnitudes of their transversal momenta,
these $n$ virtual gluons can be separated in the following
kinematic classes:
\begin{multline}
\vecc{q}^2\gg \vecc{k}_{i_m}^2\gg \vecc{k}_m^2,\quad
\vecc{k}_m^2\gg \vecc{k}_{i_{m-1}}^2\gg \vecc{k}_{m-1}^2 \ ,\\
,\dots,\quad \vecc{k}_1^2\gg \vecc{k}_{i_0}^2\gg \lambda^2 \ ,
\end{multline}
with the ordering in each class
$$\vecc{k}_{1_l}^2\gg\vecc{k}_{2_l}^2\gg\dots\gg\vecc{k}_{{n_j}_l}^2 \ ,
\quad l=0,1,\dots,m \ . $$ As a result, we obtain
\begin{equation}
{\cal M}_n^{(m)}=\sum_{n_j}\prod_{i=1}^m\frac{1}{(n_i)!}{\cal M
}^{n_i}_{(i)}= \frac{1}{n!}\(\sum_{n=0}^m {\cal M}_{(n)}\)^n \ ,
\label{sum}
\end{equation}
and the statement (\ref{eq:be}) immediately follows from Eq.
(\ref{sum}).

\section{Conclusions and outlook}

We emphasize that the quark inelastic form factor with emission of
$m$ real hard gluons cannot be expressed in terms of the elastic
form factor, in contrast to the QED case. Therefore, the Poisson
distribution is violated in QCD. Nevertheless, it can be shown
that the quantity
\begin{eqnarray}
\sum_{m=0}^\infty \frac{1}{m!}\Big|{\cal M
}_\infty^{(m)}\Big|^2\prod_{i=1}^m \(
g^2\frac{\dd^3k_i}{2\omega_i(2\pi)^3} \)
\end{eqnarray}
does not depend on the external virtuality $s$ that is, in fact,
the consequence of the KLN theorem \cite{KLN}. This statement can
be considered as a generator of relations between certain
contributions in each order of perturbative expansion. This sort
of relations in one- and two-loop orders was studied in \cite{KF}.
It is necessary to keep in mind that the decay type situation
$\a_i\b_j \approx \a_j\b_i$ plays an important role in this check
problem. In the present work, we have considered only the
kinematics in which the emitted real gluons are ordered according
to their transversal momenta
\begin{eqnarray}
\vecc k_1^2 \gg \vecc k_2^2 \dots \gg {\vecc k}_n^2 \ ,
\label{dec1}
\end{eqnarray}
which produces the DL contribution to the total cross section, but
the decay kinematics drops out in this regime. The condition
(\ref{dec1}) can be formulated in a Lorentz invariant form using
the relation
\begin{eqnarray}
\vecc k_i^2 = s \a_i \b_i = 2 \frac{k_i\cdot p_1 \ k_i \cdot
p_2}{p_1\cdot p_2} \ . \label{dec2}
\end{eqnarray}
We believe that the result for inelastic form factor, Eq.
(\ref{eq:be}), under the condition (\ref{dec1}) can in principle
be verified by relevant exclusive experiments.

It is worth noting that the arguments given above do not take into
account the nature of external particle which can be any probing
particle including scalar, pseudoscalar, vector and pseudovector
particle ($\Gamma=1;\gamma_5;\gamma_\rho;\gamma_5\gamma_\rho$). In
particular, all the results are valid for the quark Pauli form
factor. Moreover, just a small modification must be made when the
flavor of one of the quarks changes. We will not consider this
case here.

Let us emphasize that Gribov's idea on the pole-dominated
contribution of virtual exchange particle with minimal transversal
momentum can be effectively applied to any gauge theory including,
{\it e.g.}, gravitation. In the latter case, the virtual exchange
particles as well as the real emitted ones are gravitons.

We point out two possible applications of the results obtained in
the present work. One concerns the decay of a heavy particle
current described by $\Gamma$ to the quark-antiquark pair
accompanied by an arbitrary number of gluons. Here the form factor
reveals itself in the time-like region. Another possible
application is the mechanism of the jet formation in DIS
experiments where the spacelike region can be probed.

\begin{acknowledgments}
Two of us (E.B., E.A.K.) are grateful to the Institute of Physics
SAS, Bratislava, where part of this work was carried out, for the
warm hospitality. One of us (E.A.K.) is also grateful to the
Theory Departments of Saint-Petersburg and Novosibirsk Nuclear
Physics Institutes, and personally to L.N. Lipatov for valuable
fruitful discussions. We thank B.I. Ermolaev for important
remarks. The work was supported in part by RFBR (Grants Nos.
03-02-17077, 03-02-17291, 04-02-16445), Russian President's Grant
No. 1450-2003-2, and INTAS (Grant No. 00-00-366).
\end{acknowledgments}

\section*{Appendix}
In QCD, the 1-loop order of PT contribution to the elastic form
factor in the scattering channel has the form
\begin{multline}
{\cal M}^{(1-loop)} = -\frac{ig^2}{(2\pi)^4}\int\frac{\dd^4k}{\(k^2-\lambda^2+i0\)}\\
\times \frac{\bar{u}(p_2)\gamma_\mu\sigma_a(\hat{p}_2-\hat{k}+m)
\Gamma(\hat{p}_1-\hat{k}+m)\gamma_\mu\sigma_a u(p_1)} {\(
(p_2-k)^2-m^2+i0 \) \((p_1-k)^2-m^2+i0\)} \ . \label{app1}
\end{multline}
Simplifying the numerator and neglecting the power suppressed part
(related to the Pauli form factor) one finds $2sC_FV_0$ in the
numerator. Using the Sudakov parameterization of the loop momentum
\begin{align}
k^2-\lambda^2+i0&=s\a \b - \vecc{k}^2-\lambda^2 + i0,     \\ \nn
(p_1-k)^2-m^2+i0&=-s\a(1-\b)-\vecc{k}^2 + i0, \\ \nn
(p_2-k)^2-m^2+i0&=-s\b (1-\a)-\vecc{k}^2 + i0 \ ,
\end{align}
and analyzing the location of poles in the $\a$, $\b$ planes, one
finds that the nonzero DL contribution arises from two situations
corresponding to the location of the poles of the integrand in
different half-planes of $\a$-plane:
\begin{align}
0< (\b , \a) < 1\ , \ \ s\a\b > \lambda^2 \ , \ \\
0 < (- \b , - \a) < 1 \ , \ \ s\a\b > \lambda^2 \ \ .
\end{align}
Performing the $\dd^2\vecc{k}$ integration
\begin{multline}
\int\frac{\dd^2\vecc{k}}{s\a\b - \vecc{k}^2 - \lambda^2 + i0}=\\
-i\pi^2 \int \dd\vecc{k}^2\delta(s\a\b - \vecc{k}^2 - \lambda^2)=
-i\pi^2\theta(s\a\b-\lambda^2) \ ,
\end{multline}
we arrive at
\begin{eqnarray}
2\int\limits_{\frac{\lambda^2}{s}}^1\frac{\dd\a}{\a}
\int\limits_{\frac{\lambda^2}{s}}^1\frac{\dd\b}{\b}\theta(s\a\b-\lambda^2)=
\ln^2 \frac{s}{\lambda^2} \  ,
\end{eqnarray}
which immediately yields the one-loop amplitude in the double-log
approximation:
\begin{equation}
{\cal M}^{(1-loop)} = - V_0 \ C_F \frac{\a_s}{4\pi} \ln^2
\frac{s}{\lambda^2} \ . \label{qcd1}
\end{equation}
The QED result can be obtained from Eq. (\ref{qcd1}) by the
replacement $C_F \to 1$.

In order to calculate the form factor (\ref{app1}) in the case
when the transversal momentum of the emitted real gluon is taken
as an upper bound for the virtual boson transversal momentum, one
needs to take into account another $\theta$-function:
$\theta\(s\a\b - k_\bot^2 \)$ in the loop integration.  Then, one
gets
\begin{multline}
2\int\limits_{\frac{\lambda^2}{s}}^1\frac{\dd\a}{\a}
\int\limits_{\frac{\lambda^2}{s}}^1\frac{\dd\b}{\b}\[
\theta(s\a\b-\lambda^2) - \theta\(s\a\b - k_\bot^2 \) \] \\ =
\ln^2 \frac{s}{\lambda^2} - \ln^2 \frac{s}{k_\bot^2} \  ,
\end{multline}
thus obtaining the contribution of the ``down'' loops in Eq.
(\ref{em1ph}).

\end{document}